\documentclass[
 reprint,
 amsmath,amssymb,
 aps,
 prl
]{revtex4-2}

\usepackage{graphicx}
\usepackage{dcolumn}
\usepackage{bm}

\begin{document}

\preprint{APS/123-QED}

\title{Orbital-Selective Engineering of Strain-Tunable Chern Insulators in Momentum Space}
\thanks{The first three authors contribute equally to this work}

\author{Jin Gao}
\affiliation{Key Laboratory of Magnetism and Magnetic Functional Materials (Lanzhou University), Ministry of Education, Lanzhou, China}

\author{Rongrong Chen}
 \altaffiliation[Current address:]{Henan Institute of Technology, Zhengzhou, China}

\author{Lei Yang}%
\affiliation{Key Laboratory of Magnetism and Magnetic Functional Materials (Lanzhou University), Ministry of Education, Lanzhou, China}

\author{Zixiong Huang}%
\affiliation{School of Physics Science and  Technology, Lanzhou University, Lanzhou, China}

\author{Dingzhao Liu}%
\affiliation{School of Physics Science and  Technology, Lanzhou University, Lanzhou, China}

\author{ChengLong Jia}%
\email{cljia@lzu.edu.cn}
\affiliation{Key Laboratory of Magnetism and Magnetic Functional Materials (Lanzhou University), Ministry of Education, Lanzhou, China}

\author{Li Xi}
\affiliation{Key Laboratory of Magnetism and Magnetic Functional Materials (Lanzhou University), Ministry of Education, Lanzhou, China}

\author{Desheng Xue}
\affiliation{Key Laboratory of Magnetism and Magnetic Functional Materials (Lanzhou University), Ministry of Education, Lanzhou, China}

\author{Kun Tao}
\email{taokun@lzu.edu.cn}

\affiliation{Key Laboratory of Magnetism and Magnetic Functional Materials (Lanzhou University), Ministry of Education, Lanzhou, China}

\date{\today}

\begin{abstract}



Unlike conventional approaches where topological order is statically fixed, we demonstrate that biaxial strain can independently modulate topological order and functional responses in a Tc-adsorbed penta-hexa silicene monolayer. Combining first-principles calculations and tight-binding models, we establish a continuous topological pathway \(C=+1 \rightarrow 0 \rightarrow -1\) driven purely by strain, with the intermediate \(C=0\) state being a gapped insulator rather than a gapless critical point. This evolution is governed by momentum-space orbital-selective engineering, which selectively reconfigures wavefunctions to reshape the global Berry curvature distribution. Crucially, we reveal a fundamental dichotomy: topology originates from the global phase distribution of orbital hybridization, while piezoelectric functionality arises from its local strength. This enables the coexistence of nontrivial Chern insulating states with giant electromechanical responses (\(d_{11} \approx 11\) pm/V, three times that of $MoS_{2}$), establishing a paradigm for transforming static functional materials into dynamically tunable quantum platforms.

\end{abstract}

\maketitle


The discovery of topological quantum states, epitomized by the quantum anomalous Hall insulator (QAHI), has redefined the landscape of condensed matter physics\cite{Haldane, Rui-Yu, Weng, Heke, Liu, Tokura2019, Chang}. These states host dissipationless chiral edge channels protected by a global topological invariant, the Chern number (C), offering a robust platform for next-generation electronics. Characterized by a non-trivial bulk topological Chern number \( C \) , this class of materials is also known as Chern insulators\cite{Thouless}. To date, Chern insulators have been achieved in a variety of two-dimensional systems, such as magnetically doped films  $(Bi,Sb)_{2}Te_{3}$ \cite{Chang2013, Chang2015, Kou2014, Ou2018, Deng2020},  intrinsic $MnBi_{2}Te_{4}$\cite{Liu2020, Ge, Liu2021}, moiré graphene and transition metal dichalcogenide (TMD) heterostructures\cite{Serlin, Li2021}, as well as bilayer $V_{2}O_{3}$, the $V_{2}MX_{4}$ monolayer, and the $M_{2}X_{2}$ family\cite{Mellaerts2021, Duan2024, LiYang, Wang2021, LiuLei, Jiang2024, Guo2022}. The capability of dynamically controlling  both the sign and magnitude of the Chern number in the topological insulator allows precise control of the propagation direction of chiral edge channels, which is a direct manifestation of the delicate manipulation of the time-reversal symmetry breaking and band structure\cite{Ding25, Zheng2025}. This offers great value in driving innovative developments in topological electronics, spintronics, and quantum computing.  

However, the field faces a persistent control problem\cite{Zhao2020}. In most realized materials, the Chern number (C), the fundamental topological invariant, remains static after synthesis, severely limiting applications in reconfigurable quantum devices\cite{Yao}. Strain engineering is widely seen as a promising way forward, offering a clean and non-invasive way to modify the lattice\cite{Jenus, Lopes, Chen, Qiao2022, Tian}. However, much of the existing work is confined to a relatively elementary level of control: switching between the critical (C=0) and topological (C=±1) phases\cite{Jenus, Lopes, Zhang}. Although recent breakthroughs have successfully realized higher-Chern-number states in systems such as  the bilayer $V_{2}O_{3}$ or the $M_{2}X_{2}$ family\cite{Mellaerts2021, Duan2024, LiYang, Wang2021, LiuLei, Jiang2024, Guo2022}, these states are often "locked in" by specific stacking geometries or elemental compositions\cite{Sun_2022, Bhattarai, Bao_2023, Zhu, DuAo, Qiao2025}. Their topological transitions tend to be abrupt, lacking the fine-tuned continuous evolution needed for practical manipulations\cite{ Wang, Lijq, WangShu, JiShilei, Lv_2022}.

A more ambitious and technologically relevant goal remains largely unmet: the quantitative stepwise manipulation of the Chern number in a monolayer material using a single continuous external knob such as strain\cite{Yao}. Or even further, can we use a single knob to independently control what it is (topological order) and what it does (functional response) within a single material? Why is this so difficult? The central challenge lies in the coarse theoretical treatment of strain, which often describes the effect of strain in overly coarse terms as a uniform lattice scaling\cite{Mannai_2020, Sun_2022}. This fails to capture how a macroscopic deformation selectively rewrites wavefunctions at specific points in momentum space (k-space) to reshape the global Berry curvature distribution. Recent studies have begun to emphasize the importance of orbital degrees of freedom\cite{Bhattarai}, however, a predictive framework that connects macroscopic strain to the k-resolved orbital response and further to the global topological order remains lacking.  While $TbTi_{3}Bi_{4}$ exhibits static orbital selectivity \cite{Cheng2025, Zhang2025}, and $GdTi_{3}Bi_{4}$ displays spin-intertwined charge order \cite{Han2026}, actively manipulating  these phenomena with an external knob remains a challenge.

To bridge the disconnect between topology and multifunctionality, we demonstrate that a single external knob—biaxial strain—can independently modulate topological order and functional responses in Tc@PH-Si through selective orbital engineering of momentum-space $Tc-d_{xz}/Si- p_{x}$. The strain drives a complete topological pathway: C=1 (0\%) $\rightarrow$  C=0 (-2\%) $\rightarrow$  C = -1 (-3\% to -4\%) $\rightarrow$  C = 0 metallic (-6\%). Crucially, the intermediate point C=0 at -2\% strain hosts a direct band gap (0.17 eV) and a giant piezoelectric response ($d_{11}$ =8.34 pm/V), while the state C = -1 at -4\% strain achieves $d_{11}$ = 11.01 pm / V—three times that of $MoS_{2}$. Berry curvature analysis reveals that functionality arises from the local orbital hybridization strength, whereas topology originates from global phase distribution. This establishes a new paradigm: the transformation of static functional materials into dynamically tunable quantum platforms through a single, unified mechanism.

Monolayer PH-silicene is an intrinsic 2D magnetic semiconductor (P31m space group) with room-temperature stability\cite{PH-silicene}. Its magnetic order and electronic properties are tunable via strain or stacking without transition-metal doping. Substituting Si with transition metals yields PH-SiX monolayers (X = Cd, Zn, Bi, Ga, Al), which exhibits high piezoelectricity (e.g 63.1 pm V$^{-1}$ for PH-SiZn), tunable band gaps (0.11–1.07 eV) and a giant Rashba effect in PH-SiBi ($\alpha_{R}=1.18$ eVÅ$^{-1}$), promising for functional devices\cite{Yang2024}. Transition-metal adsorption creates PH-SiBiX (X = Sc–Cd), a multifunctional platform\cite{Yang2025}. It simultaneously hosts record piezoelectricity (d$_{11}=13.06$ pm V$^{-1}$ for Ti), auxeticity ($\nu \sim -0.206$ for Ni), strain-tunable visible-light absorption (7.5–16\%) and strong spin-orbit coupling ($\sim$50 meV), enabling adaptive piezotronics and flexible optoelectronics.


\begin{figure}[htbp]
  \centering
\center\includegraphics[width=8cm,clip=true]{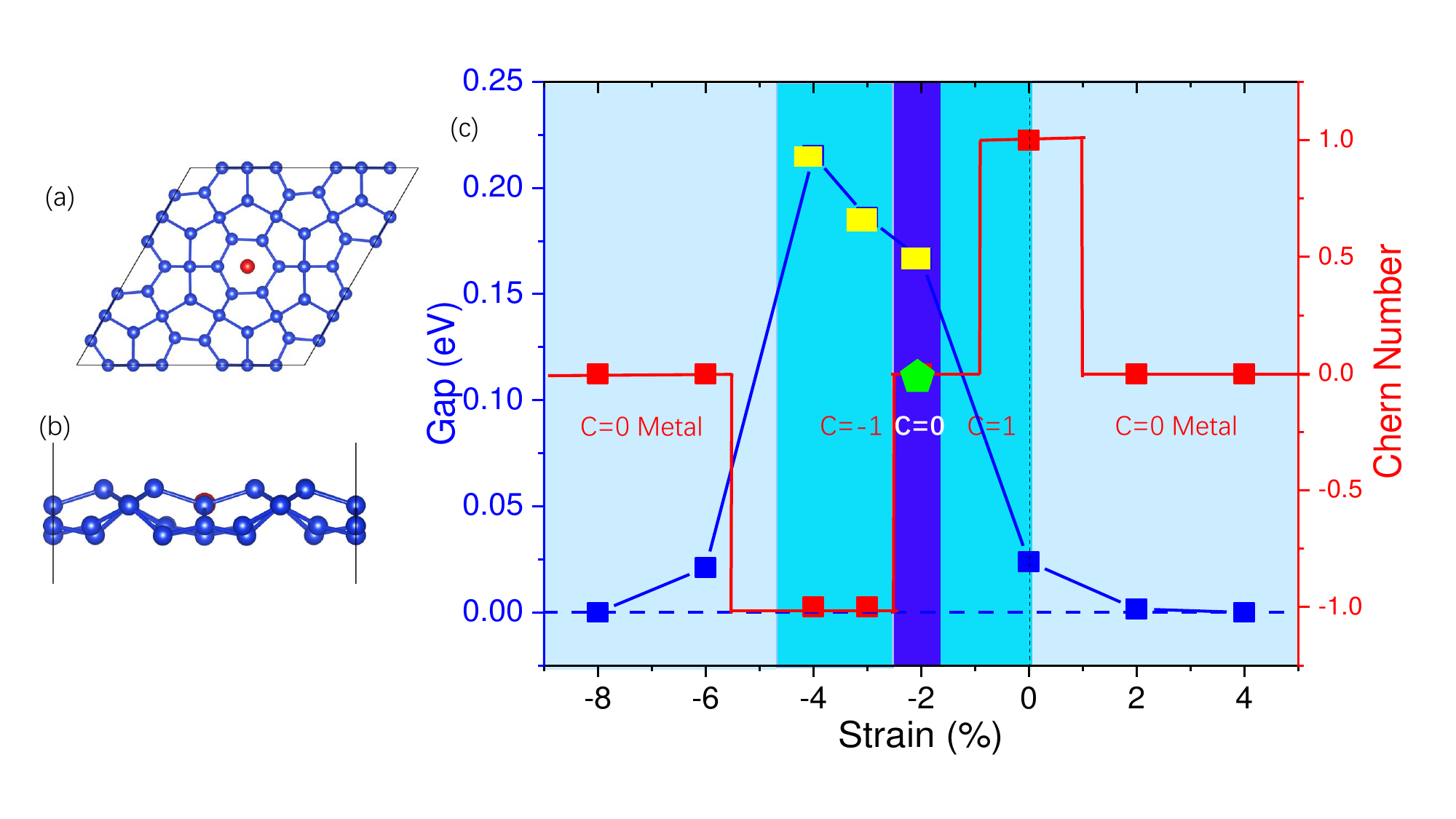}
\caption{\label{fig1}
(a) , (b) Top and side view of the Tc@PH-Si. (c) The evolution of the band gap and the Chern numbers with the strain. The yellow rectangles marked in the figure denote the direct band gap regions.}
\end{figure}


Computational details are provided in Sect. A of the \textbf{\textit{Supplementary Material}}. Our strategy for engineering a Chern insulator employs 3d and 4d transition metal atoms adsorbed on PH-silicene (denoted M@PH-Si) motivated by the progress of previous studies in graphene\cite{ding2011}.


The lattice stability and magnetic properties of all of the M@PH-Si monolayers are summarized in Tables SII and SIII. The evolution of the topological Chern number \(C\) with biaxial strain in monolayer Tc@PH-Si is summarized in Fig.  \ref{fig1} (c).  At zero strain, the system is a Chern insulator with \(C = +1\). Upon applying compressive strain, we observe a complete topological pathway: \(C = +1\) (0\%) \(\rightarrow\) \(C = 0\) (\(-2\%\)) (topologically intermediate trivial state) \(\rightarrow\) \(C = -1\) (\(-3\%\) to \(-4\%\)) \(\rightarrow\) \(C = 0\) (\(-6\%\)) (metallic state); the \(C = 0\) state in the strain \(-2\%\) is particularly noteworthy, as it marks the closure of the topological gap and the onset of a direct band gap, which remains direct throughout the strain range \(-2\%\) to \(-4\%\)  [Fig. 1(c), yellow shaded region]. The band gap reaches a maximum of \(\sim 0.22\) eV at \(-4\%\) strain before closing again at higher compression. 
This indirect-to-direct shift is critical for optoelectronic performance, as a direct band gap promotes efficient electron-hole recombination, thereby improving light absorption and emission efficiency\cite{Chorsi, Photo, Leykam2026}. 


In particular, this evolution of the band gap is closely correlated with the variation of the Chern number: the system exhibits $|C|=1$ at 0\%, -3\% and -4\% strain, respectively, while $C=0$ is observed at -2\% strain, indicating that the shift in the character of the band gap (indirect-to-direct) and the modulation of the topological invariants are synergistically regulated by strain. This strong correlation not only reveals the intrinsic link between electronic topological properties and optical performance of the material but also provides a feasible strategy for tuning the topological and optoelectronic properties of the system through strain engineering. 

\begin{figure*}[!htbp]
  \centering
  \includegraphics[width=0.95\textwidth,clip=true]{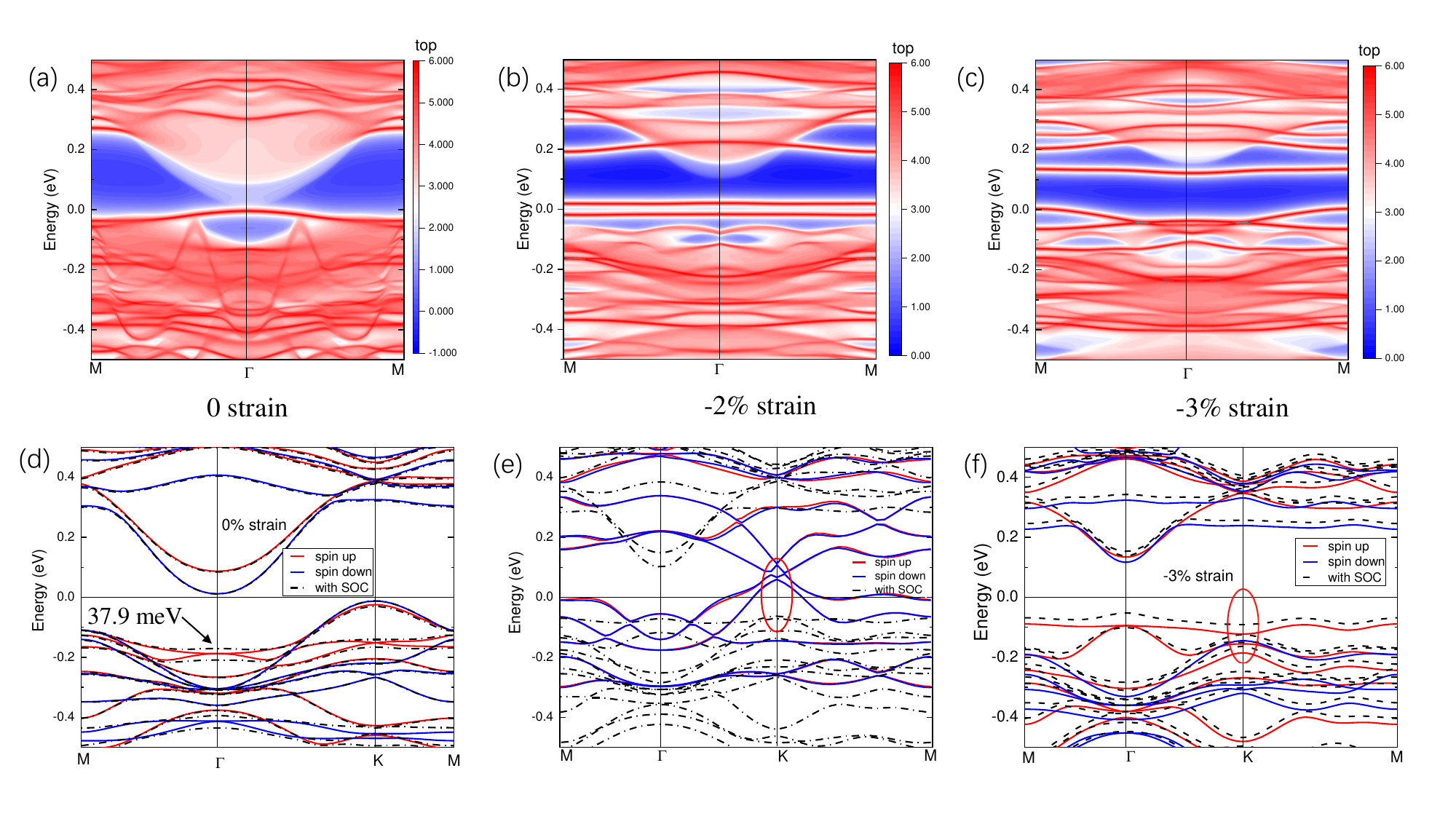}
  \caption{\label{fig2}Surface-states of the Tc@PH-Si nanoribbon under different strains: (a) 0\%, (b) -2\% and (c) -3\%. Electronic structures of Tc@PH-Si under different strains with and without SOC (d) 0\% , (e) -2\% and (f) -3\%.}
\end{figure*}

To elucidate the origin of the strain-dependent Chern numbers in Tc@PH-Si, we examined the edge states of the nanoribbons under representative strains of 0\%, -2\% and -3\%, corresponding to Chern numbers C = 1, 0 and -1 (Fig. \ref{fig2}). In 0\% stain C = 1, a single gapless chiral mode appears selectively on one surface of the buckled ribbon [Fig. \ref{fig2}(a)], originating from a band inversion at the $\Gamma$ point when spin-orbit coupling (SOC) is included [Fig. \ref{fig2}(d)]. A similar single edge state is observed for the C = -1 phase under -3\% and -4\% strain [Fig. \ref{fig2} (c)]. However, at -2\% strain, two edge-like states emerge around the Fermi level; these are not chiral surface states, as they are strongly localized and are not connected to the conduction or valence bands[Fig. \ref{fig2} (b)]. 




However, the bottom surface shows only a diffuse multi-branch structure near the Fermi level, consistent with boundary asymmetry induced by the puckered lattice; see Fig.S3. The intrinsic buckling of PH-silicene, analogous to that in buckled $\beta$-Sb monolayers, can promote higher-order topological responses\cite{Huang2021} and provides a versatile structural platform for strain-tunable Chern phases in this magnetic system. The selective emergence of a clean one-channel edge spectrum solely on the top surface is robust, and its boundary manifestation is highly sensitive to the local atomic registry—a key design principle for future topological nanodevices.

Figures 2(d)–(f) show the band structures of Tc@PH-Si at 0\%, -2\% and -3\% strain, calculated with and without spin-orbit coupling (SOC). Without SOC, the spin-up bands touch at the $\Gamma$ point at 0\% strain. Including SOC opens substantial gaps at these crossings near the Fermi level. At 0\% strain, the GGA-calculated SOC gap is 37.9 meV. The  HSE06+SOC benchmark calculations at 0\% strain confirm the topological insulating phase with a global band gap of 0.525 eV (see Fig.S1), while the topological gap governing strain-driven phase transitions remains at the 30.2 meV at the $\Gamma$ point. Although smaller than the gaps predicted in $Bi_{2}Te_{3}$\cite{Zhang2009}, $MnBi_{2}Te_{4}$\cite{Zhang2019} and $M_{2}X_{2}$\cite{ktgw-2wx2}, it is comparable to the derived gap in the TbCl
room-temperature QAHE system (~42.8 meV)\cite{Zhong}. As the compressive strain increases to -4\%, a band crossing develops along the $\Gamma$–K directions due to threefold rotational symmetry. Consequently, the SOC-induced gap at $\Gamma$ increases slightly to 45.6 meV, while the gap at K points increases to approximately 60.3 meV.

At the critical strain of \(-2\%\), where the system is at a topologically critic point (\(C=0\)) but functionally optimal, our spin-resolved projected band analysis reveals a dual orbital-level mechanism that governs the transition. The orbital Tc \(d_{z^2}\) is pulled below the Fermi level and becomes occupied in both spin channels, effectively quenching the local exchange splitting and suppressing the spin moment to \(0.012\ \mu_{\mathrm{B}}\) in spin polarized calculations, see \textbf{supplementary materials}. Concurrently, the \(d_{xz}\) and \(d_{yz}\) orbitals, which are strongly hybridized in the \(C=\pm1\) phases, become completely decoupled at the K point. This decoupling persists even when spin-orbit coupling is included (Fig.\ref{fig2}(e))—although SOC restores the magnetic moment to 2.44 $\mu_{B}$, it cannot re-establish the specific orbital entanglement required for a nonzero Chern number. These observations directly evidence the core principle of orbital-selective engineering: the strain independently manipulates both the magnetic exchange (via \(d_{z^2}\) occupation) and the topological orbital network (via \(d_{xz}/d_{yz}\) coupling), rendering \(-2\%\) a clean critical state where both prerequisites for a Chern insulator are simultaneously disabled.

\begin{figure*}[!thbp]
  \centering
  \includegraphics[width=0.90\textwidth,clip=true]{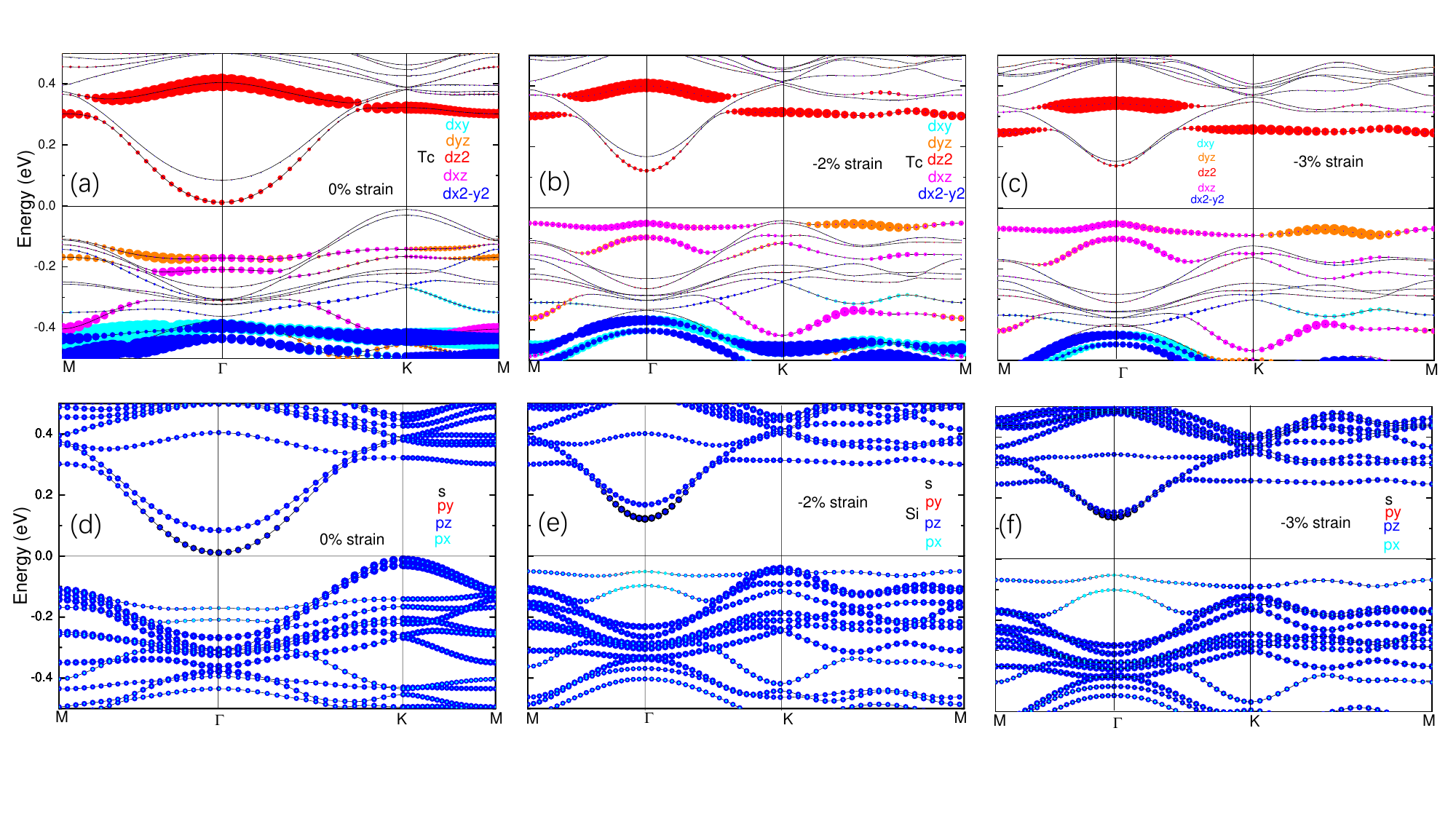}  
\caption{\label{fig3}
$d$-orbital projected band structures of the Tc in the monolayer Tc@PH-Si under different strains: (a) 0\%, (c) -2\% and (e) -3\%; Orbital projected band structures of the Si in the Tc@PH-Si under different strains: (b) 0\%, (d) -2\% and (f) -3\%. The sizes of the circles indicate the weight of different orbitals. 
}
\end{figure*}

The quantized Chern number \(C = 1\) originates from an orbital selective reconstruction mediated by spin-orbit coupling (SOC). As shown in Fig.\ref{fig2} and Fig.\ref{fig3}, SOC does not act as a uniform perturbation but rather as a symmetry-selective manipulator at the \(\Gamma\) point. It lifts the degeneracy of the Si-\(p_x\) and Si-\(p_y\) states and, more importantly, induces a pronounced splitting within the active Tc-\(d_{xz}/d_{yz}\) manifold. This splitting produces a hybrid orbital that remains degenerate with \(d_{yz}\), while shifting its counterpart to higher energy—an effect that is anisotropic in momentum space due to the inherent \(d_{xz}/d_{yz}\) energy separation along the M–\(\Gamma\) path even without SOC. This orbital selective reconstruction directly reorders band energies near the Fermi level, enabling a single band inversion. The resulting momentum-dependent hybridization 
gives rise to a spatially textured Berry curvature distribution, whose integral over the Brillouin zone is quantized to a Chern number \(C = 1\). This establishes a direct link between microscopic orbital reconstruction and the macroscopic topological invariant.

The strains $0\%$ and $-3\%$ exhibit surface states with Berry curvature concentrated at K points, the critical strain $-2\%$ reveals an orbital-decoupled intermediate state: band structure analysis shows that the Tc $d_{xz}$ and $d_{yz}$ orbitals—hybridized in both $C=\pm1$ phases—become spatially separated ($d_{xz}$ along $\Gamma$-K, $d_{yz}$ along K-M) with negligible mixing  of the K point (Fig.\ref{fig2} (b)). This creates fragmented Berry curvature multipoles that geometrically cancel, producing  C$\approx0$ despite local singularities $|\Omega_z|>10^3$ \AA $^2$ , see Fig.\ref{fig5} (d).  

We identify this as momentum-space orbital selection, where compressive strain quenches orbital entanglement without gap closing. Unlike conventional band-inversion transitions, the $C\approx0$ state is topologically critical: In particular, the WCCs display a pronounced discontinuity (jump) at ky$\sim$0.4  [Fig. S4(b)], signaling a strong hybridization between the occupied states and a rearrangement of Wannier centers, rather than a gap closing, yet a singular local Berry phase (Fig.\ref{fig5} (d)). The $d_{xz}$-$d_{yz}$ dephasing enables continuous piezo-topological switching through an off-state, opening avenues for orbital-patterned quantum devices.

\begin{figure}[htbp]
  \centering
    \center\includegraphics[width=8cm,clip=true]{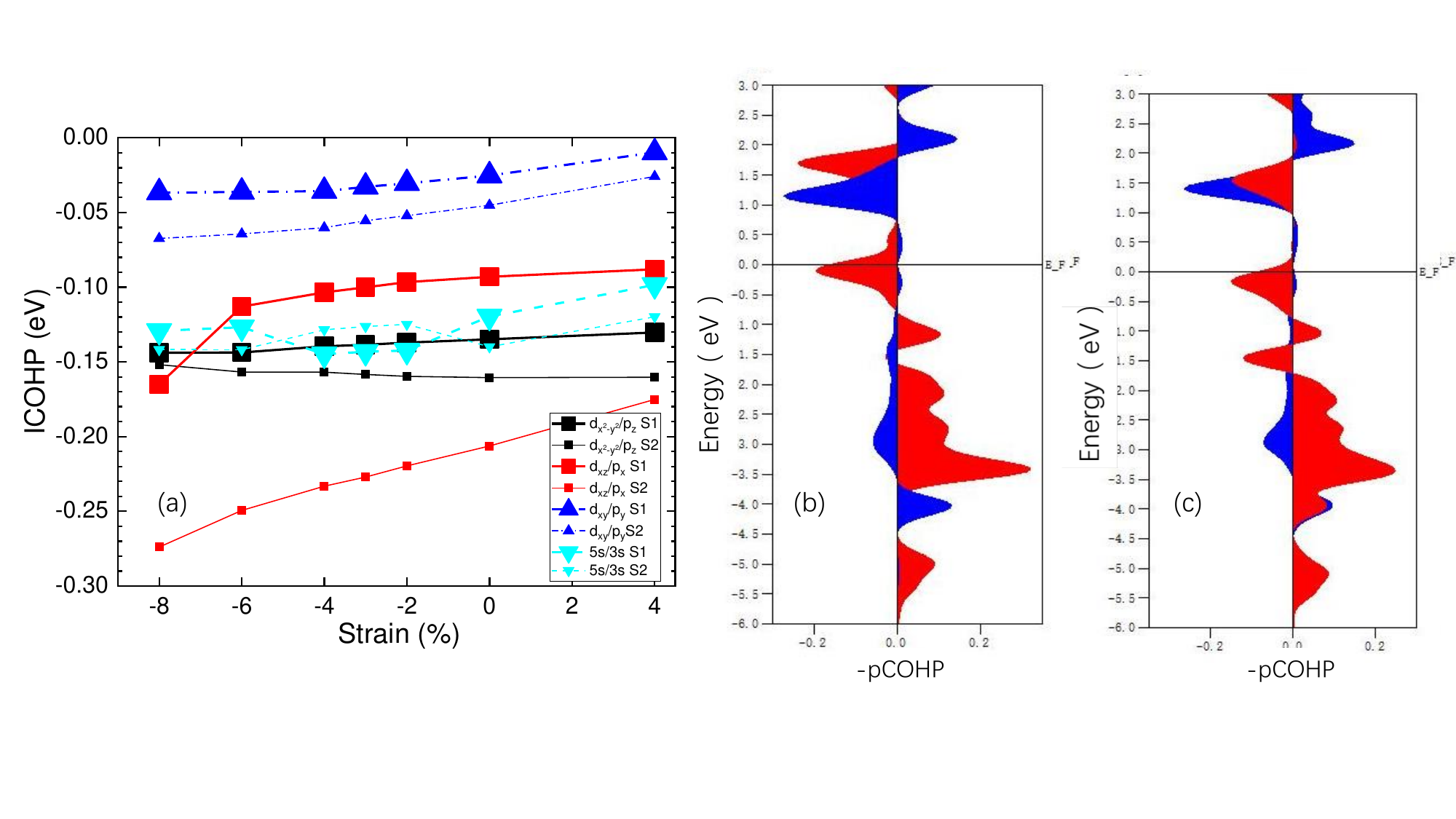}
\caption{\label{fig4}
(a) Strain dependence of ICOHP for various orbital pairs in Tc@PH-Si, in which $S_{1}$/$S_{2}$ means spin up/down, respectively. (b)Projected COHP (pCOHP) for Tc-$d_{xz}$/Si-$p_{x}$ (red) and Tc-$5s$/Si-$3s$ (blue) orbital pairs in Tc@PH-Si at $-3$\% (left) strain (C = -1) and  0\% (right) strain (C = 1).}
\end{figure}

\begin{figure}[htbp]
  \centering
    \center\includegraphics[width=8cm,clip=true]{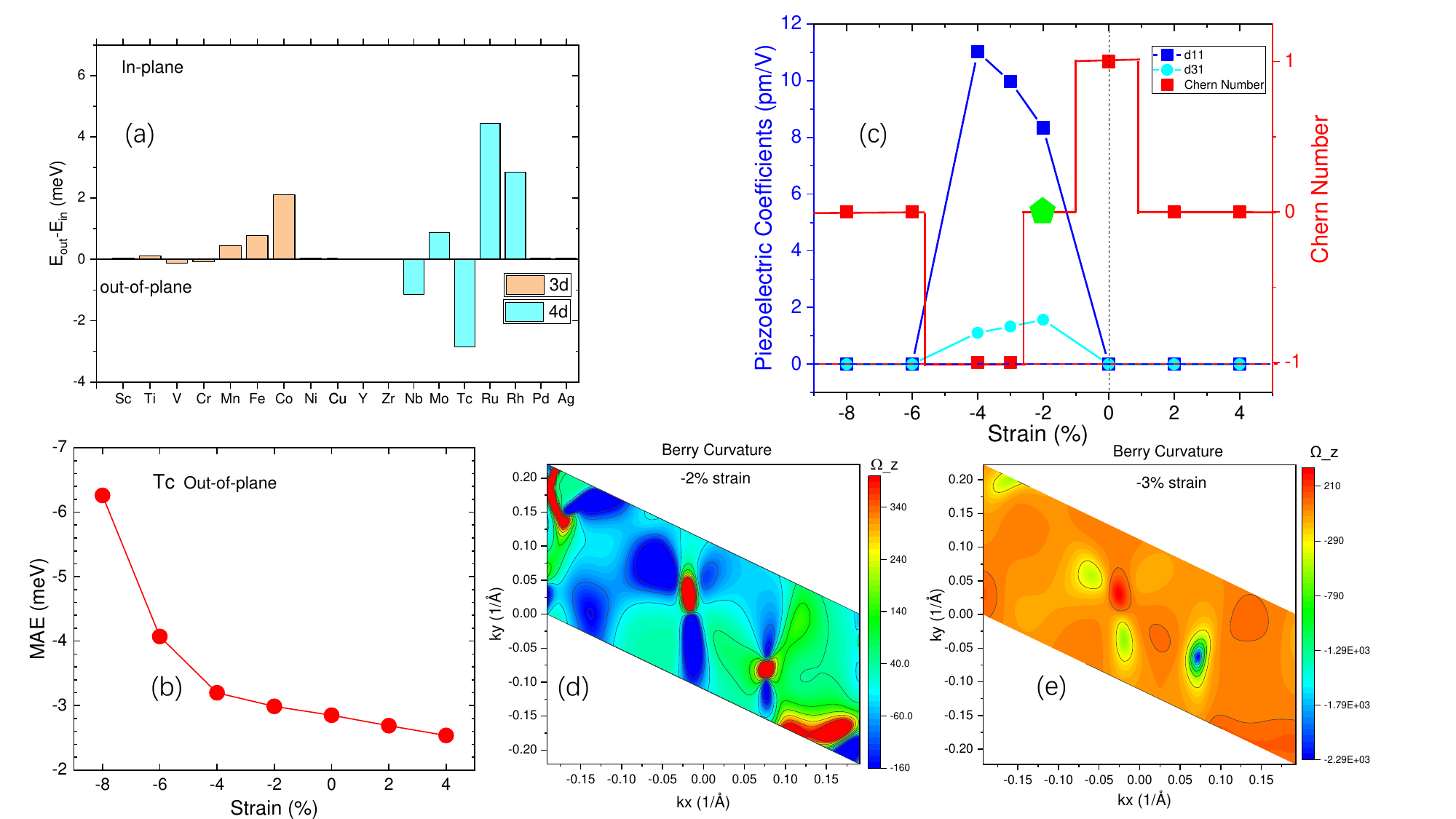}
    \caption{\label{fig5}
(a)Magnetic anisotropy energy (MAE = $E_{out}$ - $E_{in}$) for all 3d and 4d transition metals adsorbed on PH-silicene. Negative (positive) values indicate an out-of-plane (in-plane) easy axis; (b)Strain dependence of the MAE for Tc@PH-Si, with the easy axis along the out-of-plane direction; (c) Strain dependence of the Chern number (red triangles) and piezoelectric coefficients $d_{11}$ (blue squares) and $d_{31}$ (cyan circles) in Tc@PH-Si; (d) and (e) Berry curvature at 0\% and -3\% strain.}
\end{figure}

To trace the strain-driven evolution of the band structure, we analyze the orbital-projected bands in the presence of spin-orbit coupling in Fig. \ref{fig3}, which represent a weight decomposition adapted to symmetry rather than an exact eigen state. The physical reliability of evolutionary trends in orbital weights has been cross-validated through COHP bonding analysis (Fig.\ref{fig4}). This reveals that the reconfiguration of the Chern number from C = +1 to  C= -1 occurs via a precise two-stage reconfiguration of orbital interactions in momentum space. 
The transition from 0\% to –3\% strain acts as a topological reset. At the K point, the dominant character of the highest valence band shifts from Si-$p_z$ (hybridized with $p_x/p_y$) to strongly coupled Si-$p_x/p_y$ states at the $\Gamma$ point. 
corresponding to a global weakening of key hybridizations: between Si-$p_z$ and Si-$p_x/p_y$, and between the crucial Tc orbitals $d_{xz}$ and $d_{yz}$. Simultaneously, the contribution of the Tc $d_{z^2}$ orbital to the conduction-band minimum at $\Gamma$ decreases. These coordinated changes dismantle the specific orbital network that stabilized the original order C = +1. As the strain increases to \(-2\%\) (\(C = 0\)), the \(d_{xz}\) and \(d_{yz}\) orbitals decouple at the \(K\) point, leading to a redistribution of the Berry curvature that cancels the net Chern number. 
Crucially, the C=0 state at -2\% strain is not a conventional quantum intermediate trivial state (where the band gap closes), but rather a dynamic equilibrium state in which Berry curvature contributions from $\Gamma$ and K points precisely cancel while the band gap remains open (\ref{fig5}(d)), 
providing direct evidence that topological phase transitions can proceed through well-defined intermediate states rather than abrupt jumps.


The stabilization of the C = -1 phase at -3\% strain is orchestrated through a synergistic two-front operation in momentum space. First, at the K point, a new strong coupling emerges between the previously isolated Tc-$d_{xz}$/$d_{yz}$ orbitals, establishing a new channel for band inversion and Berry curvature generation. Second, at the $\Gamma$ point, the coupling between the Tc-$d_{xy}$/$d_{x^{2}-y^{2}}$ orbitals is dramatically enhanced. 
These changes are further corroborated by a pronounced up-shift ($\sim$ 0.2 eV) of the Tc-$d_{z^{2}}$ orbital and a purification of the Si-$p_{z}$ character at the edge of the conduction-band , collectively enhancing Berry curvature (Fig.\ref{fig5}(e))while maintaining C = -1. 


In essence, biaxial strain functions act as a quantum manipulator in momentum-space . 
It mirrors the orbital selective physics recently identified in monolayer M$_2$X$_2$ systems\cite{ktgw-2wx2}, where the Chern number is determined not by the count of inverted bands at a single k point, but by the symmetry-protected spatial distribution of crossings across the Brillouin zone. The strain-driven orbital engineering in Tc@PH-Si resonates with the orbital selectivity in $TbTi_{3}Bi_{4}$\cite{Cheng2025}, but elevates it from a static property to a dynamic tool. Its multifunctional enhancement mirrors the intertwined orders in $GdTi_{3}Bi_{4}$\cite{Han2026}, yet is orchestrated by a single stimulus.


Crystal orbital Hamiltonian population (COHP) analysis directly links microscopic bonding to macroscopic topology, revealing that the strain-driven topological transition results from a highly selective and spin-dependent reconstruction of specific chemical bonds\cite{COHP, Deringer2011, lobster}.

The strain-dependent integrated COHP (ICOHP) in Fig. \ref{fig4} quantifies this evolution of the bonding  in Tc@PH-Si. Under compression (–4\% to –8\%), the ICOHP for the pair Tc-$d_{xz}$/Si-$p_{x}$ (red squares) becomes substantially more negative, signaling a marked strengthening of the covalent bonding—a key driver for stabilization of phase C = -1 at -3\%. In contrast, the pairs $d_{x^2-y^2}$/$p_z$ (black squares) and $d_{xy}$/$p_y$ (blue triangles) show only minor changes, while the s/s interaction (cyan triangles) remains nearly constant. This orbital-specific response confirms that the topology is governed by directional hybridization, Tc‑$d$ and Si‑$p$ with the channel $d_{xz}$/$p_x$ playing the dominant role. The non-monotonic strengthening of the compression bonding in $d_{xz}$/$p_x$ is directly correlated with the transition from a single‑k‑point (C = 1) to a multi‑k ‑ point (C = -1) inversion of the band, thus establishing a quantitative connection between the bonding and the topological order at the microscopic to macroscopically level.



At –3\% biaxial strain, a strain-induced enhancement of orbital hybridization drives Tc@PH-Si into a C = -1 Chern insulator. The projected COHP (pCOHP) analysis in Fig. \ref{fig4}(b) identifies the interaction Tc–$d_{xz}$/Si–$p_x$ as the key driver: under compression, its bonding character (negative pCOHP) intensifies near $E_F$, while antibonding states shift to higher energies. This strengthened covalent bonding along the x-direction directly modifies band dispersions near high-symmetry points—particularly along $\Gamma$–K and $\Gamma$–M—promoting multi-k point band inversions. By contrast, at 0\% strain the weaker Tc–$d_{xz}$/Si–$p_x$ coupling confines the inversion to $\Gamma$, resulting in C=1.

Unlike previous studies explaining why different materials exhibit static Chern numbers (C=1 or 2)\cite{ktgw-2wx2}, this work demonstrates how to engineer a dynamic topological pathway (C = 1$\rightarrow$ 0$\rightarrow$-1) within a single material using strain as a continuous external knob. The pathway is governed by a resolved sequential reconfiguration of orbital couplings and establishes a direct quantitative bridge between orbital-specific bonding and macroscopic topological order. 

This orbital selective bond response is further confirmed by the spin-antiparallel, non-monotonic evolution of the semi-core Si‑3s/Tc‑5s interaction (Fig. \ref{fig4}a). Within the 0\% to –4\% strain range, the ICOHP strengthens for spin‑up electrons but weakens for spin‑down electrons. This spin-dichotomy signals a profound, global spin-polarized electronic reconstruction. The observation that even a semi-core s‑s interaction far from the Fermi level exhibits such spin-polarized changes underscores that the strain-induced  transition C = 1 $\rightarrow$ C = -1 involves a global redistribution of spin density across the entire valence manifold, extending well beyond a mere band inversion at the Fermi level. Thus, strain acts as a precision tool for tailoring topology by selectively modulating the strength of key bonds that govern the effective low-energy Hamiltonian. Variations in the lattice constant that modulate the lengths and angles of the Tc-Si bonds, thus quantitatively tuning the specific orbital overlaps of $d-p$. 

At –4\% strain, where the system enters Chern phase C = -1, Tc@PH-Si exhibits a large in plane piezoelectric coefficient \(d_{11} \sim 11.5\ \text{pm/V}\) (Fig. \ref{fig5}). This value is approximately three times higher  than that of the monolayer \(2H\text{-}MoS_2\) (3.654 pm/V)\cite{Duerloo, Wu2014}—a reference 2D piezoelectric—and significantly exceeds conventional bulk piezoelectrics such as GaN (3.1 pm / V), \(\alpha\)-quartz (2.3 pm / V) and AlN (5.1 pm / V)\cite{Bechmann, AlN, Lueng, Hangleiter}. The dramatic enhancement coincides precisely with the topological transition, revealing deep correlations between the orbital-driven band inversion and the lattice-mediated electromechanical response. The compressive strain improves hybridization between the Tc-\(d_{xz}\) and Si-\(p_x\) orbitals, which simultaneously drives multi-k-point band inversions (along \(\Gamma\text{–}K\) and \(\Gamma\text{–}M\)) that stabilize the C = -1 phase and promotes charge redistribution that amplifies piezoelectric polarization.

In  particular, the system also shows a robust out-of-plane piezoelectric response (\(d_{31} \sim 1.1\ \text{pm/V}\) at –4\% strain), exceeding functionalized h-BN (0.130 pm/V)\cite{Noor} and \(\alpha\text{-}In_2Se_3\) (0.415 pm / V)\cite{Hu2017}.This coexistence of strong in-plane and out-of-plane piezoelectricity—intrinsically linked to the buckled geometry and magnetic topology—establishes Tc@PH-Si as a unique platform for strain-tunable topological piezoelectric devices, in which quantum transport and mechanical-to-electrical conversion are unified under a single external control. Our work differs from Yu’s established approach\cite{Yu2020}, where piezoelectricity is used as a probe for topological phase transitions ($Z_2$) in time-reversal invariant systems, while our work actively engineers piezoelectricity as a tunable functional property in magnetic systems.

We also mapped the magnetic anisotropy energy (MAE) for all 3d and 4d transition metals adsorbed on penta‑hexa silicene; see the \textbf{supplementary material}. MAE shows a pronounced dependence on both the d‑electron count and the adsorbate period. Among all elements, Tc exhibits the largest MAE out of plane ($\sim$–3.5 meV/Tc-atom), establishing it as the optimal candidate to stabilize perpendicular magnetism on this 2D platform. In general, the 4d series (Y–Ag) displays stronger MAE than the 3d series (Sc–Cu), with Tc showing particularly robust out‑of‑plane anisotropy. 


The MAE of Tc@PH-Si depends strongly and monotonically on the biaxial strain, Fig.\ref{fig5} (b), with the out‑of‑plane easy axis preserved throughout. Under compression, the MAE becomes increasingly negative, reaching -6.3 meV at -8\% strain—a dramatic enhancement of perpendicular magnetic anisotropy that originates from strain ‐ induced modulation of the Tc d‑orbitals (see Fig.s5). 

In summary, we have answered the central question posed at the outset: a single external knob—strain—can independently control both the intrinsic topological order and the functional responses within a single material, monolayer Tc@PH-Si. This capability is enabled by momentum-space orbital-selective engineering, where the strain acts as a k-space quantum editor that selectively manipulates the hybridization of $Tc-d_{xz}/Si-p_{x}$. Berry curvature analysis reveals the underlying mechanism: functionality arises from the local hybridization strength of the orbital, while topology originates from its global phase distribution—a principle that governs the entire evolution driven by strain. Although topologically trivial materials are known for rich functionalities (e.g., GaN for optoelectronics, $BaTiO_{3}$ for piezoelectricity), the novelty of our work is to elucidate how a single external stimulus dynamically switches topological/critical states while preserving and even enhancing functional properties. The orbital-selective engineering paradigm established here transcends specific material systems, providing a general framework for designing dynamically tunable quantum platforms where topology and functionality can be independently optimized through a single external stimulus.

\section{Data  availability}
The data supporting the findings of this article are not publicly available after publication because it is not technically feasible and/or the cost of preparing, depositing, and hosting the data would be prohibitive within the terms of this research project. The data are available from the authors upon reasonable request.

\bibliography{apssamp}

\end{document}